\documentclass[%
preprint,
 floatfix,
longbibliography,
amsmath,amssymb,
aps,
pre,
]
{revtex4-1}

\usepackage{graphicx}
\usepackage{dcolumn}
\usepackage{bm}
\usepackage[colorlinks=true, citecolor=red, linkcolor=blue,urlcolor=blue]{hyperref}
\usepackage{soul,xcolor}

\usepackage{lmodern,pdftexcmds,stmaryrd,siunitx,xcolor}
\DeclareMathAlphabet{\mathsfbi}{OT1}{\sfdefault}{bx}{sl}
\DeclareMathVersion{sfletters}
\SetSymbolFont{letters}{sfletters}{OML}{ntxsfmi}{b}{it}
\def\v{\vspace{1cm}}

\newcommand{\bcdot}{\bm{\cdot}}
\newcommand{\bnabla}{\bm{\nabla}}
\newcommand{\btimes}{\bm{\times}}

\usepackage{scalerel}
\usepackage{tikz}
\usetikzlibrary{svg.path}

\definecolor{orcidlogocol}{HTML}{A6CE39}
\tikzset{
  orcidlogo/.pic={
    \fill[orcidlogocol] svg{M256,128c0,70.7-57.3,128-128,128C57.3,256,0,198.7,0,128C0,57.3,57.3,0,128,0C198.7,0,256,57.3,256,128z};
    \fill[white] svg{M86.3,186.2H70.9V79.1h15.4v48.4V186.2z}
                 svg{M108.9,79.1h41.6c39.6,0,57,28.3,57,53.6c0,27.5-21.5,53.6-56.8,53.6h-41.8V79.1z M124.3,172.4h24.5c34.9,0,42.9-26.5,42.9-39.7c0-21.5-13.7-39.7-43.7-39.7h-23.7V172.4z}
                 svg{M88.7,56.8c0,5.5-4.5,10.1-10.1,10.1c-5.6,0-10.1-4.6-10.1-10.1c0-5.6,4.5-10.1,10.1-10.1C84.2,46.7,88.7,51.3,88.7,56.8z};
  }
}

\newcommand\orcidicon[1]{\href{https://orcid.org/#1}{\mbox{\scalerel*{
\begin{tikzpicture}[yscale=-1,transform shape]
\pic{orcidlogo};
\end{tikzpicture}
}{|}}}}

\usepackage{hyperref} 

\begin{document}
\def\v{\vspace{1.5cm}}

\setstcolor{red}


\title{Flow field disturbance due to point viscosity variations in a heterogeneous fluid}

\author{Debasish Das}
\email{debasish.das@strath.ac.uk}
\affiliation{%
Department of Mathematics and Statistics, University of Strathclyde, 26 Richmond St, Glasgow G1 1XH, Scotland, UK}%

\date{\today}

\begin{abstract}

We derive the flow field disturbance produced by point viscosity variations in a heterogeneous fluid when subject to a background flow while neglecting fluid inertia. The disturbance flow field is found to be identical to that generated by a force-dipole called stresslet. Using a combination of theory and numerical simulations, we show how the hydrodynamics of an active rigid particle is altered due to the presence of point viscosity variations, and how this can be exploited to manipulate and steer them in microfluidic environments.

\end{abstract}

\pacs{Valid PACS appear here}
\maketitle

Fluids encountered in nature and industry are usually heterogeneous. For example, blood is composed of plasma, red and white blood cells, and platelets. The plasma itself is heterogeneous as it is an aqueous solution containing organic molecules, proteins, and salts \citep{chien1975}. Similarly, interstitial fluid in solid tumours is a highly disordered environment when compared to normal tissues and this has significant consequences on nanomedicine delivery \citep{heldin2004,jain2010}. Another example in biology is the cytoplasmic matrix which is mostly an aqueous environment but made heterogeneous due to the presence of various macromolecules. More recently, it has been discovered that cells contain numerous membraneless compartments that exhibit liquid like behaviour \cite{brangwynne2009,brangwynne2011}. Examples include nucleolus and Cajal bodies in the nucleus, and P-bodies, stress and germ granules in the cytoplasm \cite{hyman2014}. It has also been reported that P-bodies dispersed within the cytoplasm have much higher viscosity, $\sim \SI{1.0}{\pascal\second}$, than their surroundings. Hence, naturally the question arises: how do we model such heterogeneous fluid environments encountered frequently in biology?

Heterogeneous fluids are abundantly found in various industrial settings as well. Many manufacturing process involve transportation and filling of polymeric materials in channels \citep{leal1979}. These processes usually require the fluid to remain homogenous at all times but deviations occur due to impurities, segregation of different polymeric constituents or geometrical imperfections in the transportation channel.  Non-uniformities in suspensions may also occur due to physical phenomena like the well-studied lateral migration of spheres in a Couette or Poiseuille flow of a viscoelastic fluid \cite{gauthier1971}. As a result, the fluid viscosity becomes spatially heterogeneous in all these cases.
\begin{figure}
    \centering
    \includegraphics[width=0.5\linewidth]{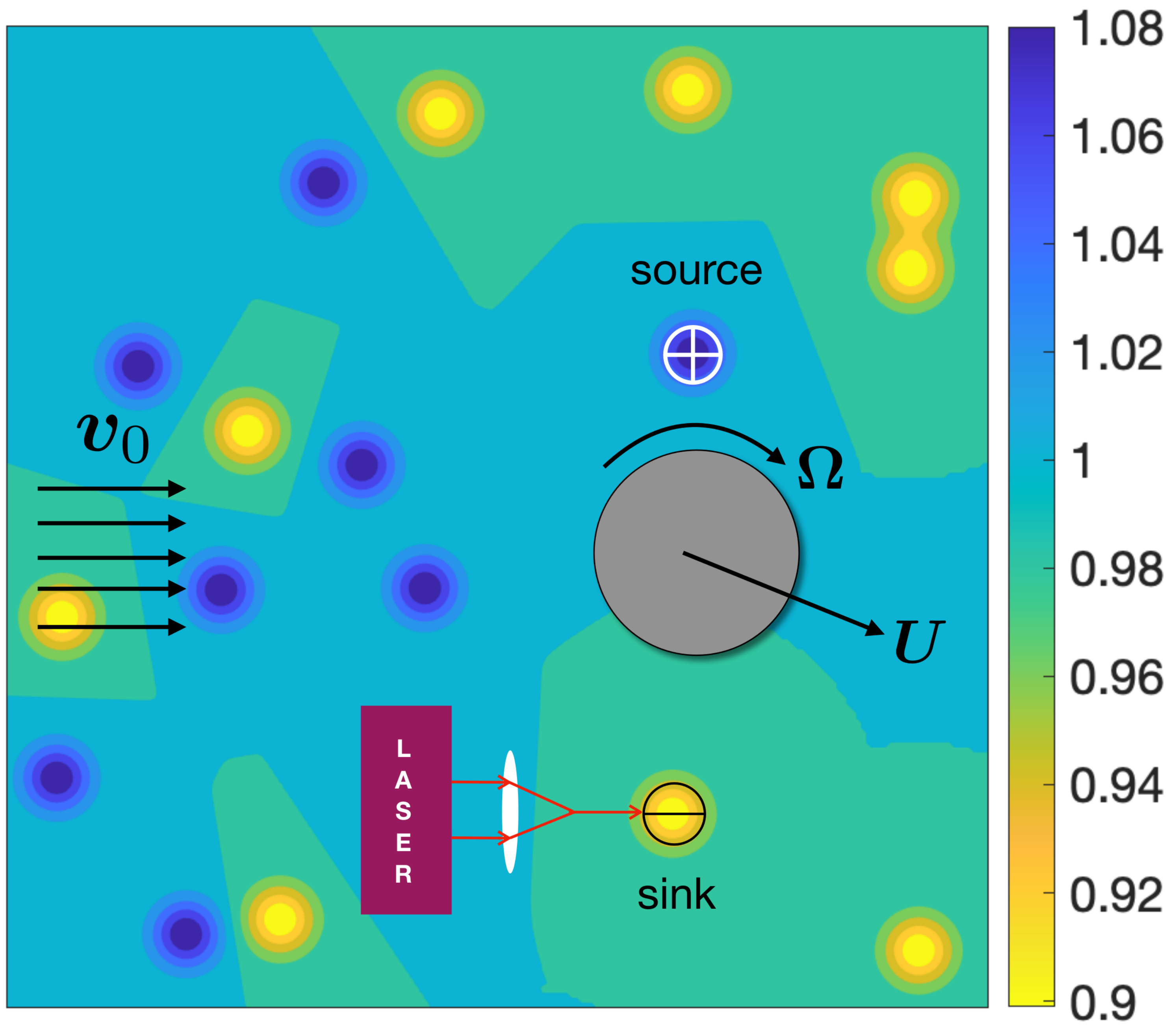}
    \caption{An illustrative example of a heterogeneous fluid with arbitrarily varying viscosity in space, $\mu (\bm{x})$, modelled as discrete viscosity sources (blue, $\oplus$) and sinks (yellow, $\ominus$) of appropriate strengths interacting with a background flow, $\bm{v}_0$. Viscosity sinks or sources may also be created in an otherwise homogenous fluid by locally heating or cooling, respectively. An active rigid particle is shown whose motion is significantly altered due to the presence of point-viscosities as its translational, $\bm{U}$, and rotational, $\bm{\Omega}$, velocity becomes coupled.}
    \label{fig1}
\end{figure}

A fluid can be heterogeneous in either its density or viscosity---its two basic material properties.  In this Letter, we focus on viscosity variations of the fluid while assuming that its density remains unchanged in space and time. A few papers have considered the effect of variable viscosity on flows in parallel \citep{pearson1977,ockendon1977,ockendon1979}, converging and diverging channels \citep{hooper1982}, and on the motion of a hot sphere \citep{morris1982,oppenheimer2016}. 
However, a fundamental fluid mechanical question has remain unaddressed: how is an ambient flow field disturbed due to a point viscosity variation in the fluid? If we are able to answer this question, we can create an arbitrary spatial distribution of viscosity discretely by placing viscosity sources and sinks of appropriate strength, allowing us to model many heterogeneous fluids such as those described above. Fig.~\ref{fig1} shows an illustrative diagram of a typical problem where multiple point viscosities interact with a background flow and consequently alter the motion of a rigid particle. Indeed, in situations where an otherwise homogenous fluid is locally heated or cooled, for example by the use of laser or spray-freezing, the model of point viscosities is directly applicable. It is noteworthy that the idea of locally heating fluid has been recently used to artificially create cytoplasmic flows inside \textit{C. elegans} zygotes \cite{mittasch2018}. The point viscosity model is valid with the assumption that the time scales of interest are smaller than the time scales at which these point viscosities diffuse, making the problem quasi-static. While it is possible to incorporate diffusion of these point viscosities in the model, we do not consider this additional complexity here. 

We restrict ourselves to an inertialess fluid whose dynamics are given by the Stokes equation \citep{lamb1932,happel2012,kim2013},
\begin{align}\label{eq:fullstokes}
\begin{split}
&-\bnabla p(\bm{x}) + \bnabla \bcdot [\mu (\bm{x}) \{ \bnabla \bm{v}(\bm{x}) + \bnabla \bm{v}^T(\bm{x}) \} ] + \bm{F} \delta(\bm{x} - \bm{x}_0) = \bm{0},
\end{split}
\end{align}
together with the incompressibility condition, $\bnabla \bcdot \bm{v} = 0$. Here, $p,\bm{v}$ are the fluid pressure and velocity that need to be determined for a given spatial distribution of viscosity, $\mu (\bm{x})$, and a point force, $\bm{F}$, acting at $\bm{x} = \bm{x}_0$.  The base flow, $\bm{v}_0$, is generated by a point force or Stokeslet, however, it may include any background flow as well, for e.g. a linearly varying flow field.
We then prescribe the fluid to have a uniform viscosity, $\mu_0$, everywhere except at certain locations, $\bm{x} = \bm{x}_\alpha$, where it has the value $\mu_0 + \mu_\alpha>0$, so that $\mu (\bm{x}) = \mu_0 + \sum_{\alpha=1}^N \mu_\alpha \delta (\bm{x}-\bm{x}_\alpha)$, $\delta$ being the Dirac delta function~\citep{lighthill1958} and $N$ being the total number of point viscosities in space. Hence, for an arbitrarily varying viscosity in space, $\mu (\bm{x})$, regions that have higher or lower viscosities than the ambient fluid are represented as concentrated viscosity sources ($\mu_\alpha>0$) or sinks ($\mu_\alpha<0$) of appropriate strengths, respectively, as a first approximation. This circumvents the use of computationally expensive volume-discretising numerical simulations and gives physical insight into the effect of the spatial variations in viscosity on the flow pressure and velocity. 

The flow field due to a point viscosity has physical meaning everywhere except at the point where they are present. 
The governing equation, Eq.~\eqref{eq:fullstokes}, takes the form,
\begin{align}
\begin{split}
&-\bnabla p(\bm{x}) +  \sum_{\alpha=1}^N  \mu_\alpha  \bnabla \delta (\bm{x}-\bm{x}_\alpha) \bcdot [ \bnabla \bm{v}(\bm{x}) + \bnabla \bm{v}^T(\bm{x}) ] \\
& + [\mu_0 + \sum_{\alpha=1}^N \mu_\alpha \delta  (\bm{x}-\bm{x}_\alpha)] \nabla^2  \bm{v}(\bm{x}) +  \bm{F} \delta(\bm{x} - \bm{x}_0) = \bm{0},
\end{split}
\end{align}
and is applicable everywhere except at $\bm{x} = \bm{x}_\alpha,\bm{x}_0$. We use Fourier Transforms to solve this equation by defining, $(p,\bm{v}) = (2\pi)^{-3/2} \int_{-\infty}^\infty (\hat{p},\hat{\bm{v}})   ~ \exp{ (i\bm{k}\bcdot\bm{x})} ~\mathrm{d}\bm{k}$, where $\hat{p},~\hat{\bm{v}}$ are the Fourier Transforms of $p,~v$, respectively.
 After some algebraic manipulations (see the appendix \ref{sec:appendixA}) for details), we find the corresponding Stokes equation in the Fourier space,
\begin{align}\label{eq:fourierspace}
\begin{split}
& - i \bm{k} \hat{p} - \mu_0 k^2  \hat{\bm{v}}  + \hat{\bm{F}}  +  (2\pi)^{-3/2} \sum_{\alpha=1}^N \big[\mu_\alpha \mathrm{e}^{-i\bm{k}\bcdot\bm{x}_\alpha}  i \bm{k} \bcdot [ \bnabla \tilde{\bm{v}}(\bm{x_\alpha}) + \bnabla \tilde{\bm{v}}^T(\bm{x_\alpha}) ] \big]= \bm{0},
\end{split}
\end{align}
together with $i \bm{k} \bcdot \hat{\bm{v}}= 0$. Crucially, we note that we have used the derivative shifting property of delta function to find Eq.~\eqref{eq:fourierspace}, see Lighthill's monograph~\citep{lighthill1958} for details. The Fourier transform of the point force is $\hat{\bm{F}} = (2\pi)^{-3/2} \bm{F}  ~ \exp{ (- i\bm{k}\bcdot \bm{x}_0)}$. Note that the velocity to be evaluated at $\bm{x} = \bm{x}_\alpha$ is independent of $\bm{k}$ and includes the velocity contribution from the point force and all point viscosities except the one at $\bm{x} = \bm{x}_\alpha$.  To make this distinction, we have replaced $\bm{v}$ with $\tilde{\bm{v}}$. Also, $\bnabla \tilde{\bm{v}}(\bm{x}) + \bnabla \tilde{\bm{v}}^T(\bm{x}) = 2 \bm{E}_\alpha$ is twice the straining flow at the location of the point viscosity at $\bm{x} = \bm{x}_\alpha$.  Taking a dot product of Eq.~\eqref{eq:fourierspace} with $\bm{k}$ eliminates $\hat{\bm{v}}$, and we obtain an equation for $\hat{p}$.
The pressure field is found by taking the inverse Fourier Transform of $\hat{p}$,
\begin{align}\label{eq:pressurefield}
p = p_0 - \sum_{\alpha=1}^N \frac{3 \mu_\alpha}{2\pi r_\alpha^5} [\bm{r}_\alpha\bm{r}_\alpha \bm{:} (\bnabla \tilde{\bm{v}})_{\bm{x} = \bm{x}_\alpha}],
\end{align}
where $p_0=\bm{F}  \bcdot \bm{r}_0/4 \pi r_0^3$ is the pressure field due to a Stokeslet, $\bm{r}_0 = \bm{x} - \bm{x}_0$, and ${\bm{r}_\alpha = \bm{x} - \bm{x}_\alpha}$. Following a similar procedure, we find the flow field due to a point force, perturbed by point viscosities,
\begin{align}\label{eq:flowfield}
\bm{v} = \bm{v}_0 - \sum_{\alpha=1}^N \frac{3\mu_\alpha}{4\pi\mu_0 r_\alpha^5}  \bm{r}_\alpha[\bm{r}_\alpha\bm{r}_\alpha \bm{:} (\bnabla \tilde{\bm{v}})_{\bm{x} = \bm{x}_\alpha}].
\end{align}
The second term on the left-hand sides of Eqs.~\eqref{eq:pressurefield} and \eqref{eq:flowfield} are the disturbance fields arising from a point viscosity, identical to those arising from a stresslet \citep{happel2012,kim2013}. The forcing for Eqs.~\eqref{eq:pressurefield} and \eqref{eq:flowfield} is the base velocity generated by a Stokeslet $
\bm{v}_0 = \bm{F}\bcdot \mathsfbi{G}/8\pi\mu_0$ where $\mathsfbi{G} = \bm{I}/{r_0} +  \bm{r}_0 \bm{r}_0/r_0^3$ is the free-space Green's function (also called Oseen-Burgers tensor). It is instructive to note that the  stresslet due to the point viscosity only acts upon the straining part of the flow field, $\tilde{\bm{v}}$.  For a single isolated point viscosity,  Eqs.~\eqref{eq:pressurefield} and \eqref{eq:flowfield} are easily solved as $\tilde{\bm{v}} = \bm{v}_0$. However, for multiple point viscosities, the equation for $\bm{v}$ becomes implicit as the point viscosities interact with each other. To capture these hydrodynamic interactions, a coupled system of equations need to be solved numerically or method of reflections may be used to make analytical progress (see the appendix \ref{sec:appendixB} for an illustrative example). Henceforth, when considering multiple point viscosities, we only retain the leading order effect and neglect hydrodynamic interactions between point viscosities, i.e. $\tilde{\bm{v}} = \bm{v}_0$. This is a valid assumption when $\mu_\alpha/\mu_0r_\alpha^3 \ll 1$, i.e. either the point viscosity variations are small in magnitude compared with the background viscosity and/or they are well separated from each other.

\begin{figure*}[t]
    \centering
    \includegraphics[width=0.95\linewidth]{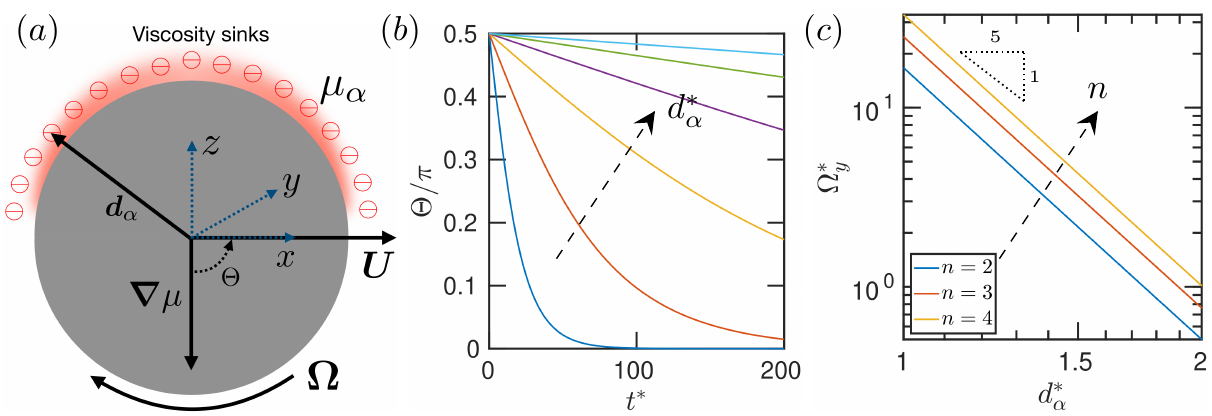}
    \caption{$(a)$ Schematic diagram of a spherical active particle surrounded by identical viscosity sinks. The viscosity sinks may be generated due to the particle surface being hot, direct heating of the fluid by a laser or due to secretions from an organism that reduce the viscosity of the surrounding medium. $(b)$ Angle made by the translational velocity of a spherical particle placed next to  viscosity sinks at various distances $d_\alpha^*=d_\alpha/a=1-2$ and the effective viscosity gradient, $\Theta$, plotted as a function of dimensionless time, $t^*=tU/a$. All the viscosity sinks are identical of strength $\mu_\alpha = -0.01\mu_0$. Their number is kept fixed at 57 and their number density, $n$, ranges from $1.1-4.5$ as $d_\alpha^*$ varies. (c) Log-log plot of dimensionless angular velocity, $\Omega_y^* = \Omega_ya/ U$, as a function of dimensionless distance between the viscosity sink and sphere's centre, $d_\alpha^*$, for various number densities, $n$.}
    \label{fig2}
\end{figure*}
Next, we consider the canonical case of a translating spherical active particle whose motion is altered due to the presence of a viscosity sink or a source. We assume that the particle has a self-propulsion velocity generated by an internal active mechanism and is torque-free. For example, it could be a squirmer \cite{lighthill1952,blake1971} or a phoretic particle \cite{howse2007}. We place a viscosity sink(s) next to one half of the translating sphere as shown in Fig.~\ref{fig2}a. The flow field due to a translating sphere of radius $a$ located at $\bm{x}_c$, 
\begin{align}\label{eq:translatingsphere}
\begin{split}
\bm{v}_0 &= \frac{3a}{4} \bm{U} \bcdot \left(\frac{\bm{I}}{r_c} + \frac{\bm{r}_c \bm{r}_c}{r_c^3} \right) +\frac{3a^3}{4} \bm{U} \bcdot \left(\frac{\bm{I}}{3r_0^3} - \frac{\bm{r}_c\bm{r}_c}{r_c^5} \right) ,
\end{split}
\end{align}
serves as the base flow, where $\bm{r}_c = \bm{x} - \bm{x}_c$. In equation \eqref{eq:translatingsphere}, we have only retained the leading order velocity flow field and neglected $\mathcal{O}(\mu_\alpha/\mu_0)$ contributions. The first and second terms are associated with flows created by a Stokeslet and source-dipole, respectively. The flow due to the translating sphere interacts with the point viscosity sink and creates a disturbance flow around the sphere itself given by the second term in Eq.~\eqref{eq:flowfield}, i.e.~$
\bm{v}_d = - (3\mu_\alpha/4\pi\mu_0 r_\alpha^5)\bm{r}_\alpha[\bm{r}_\alpha\bm{r}_\alpha \bm{:} \bnabla \bm{v}_0(\bm{x}_\alpha)]$. The vorticity due to this disturbance flow is $\bm{\omega} = [\bnabla \btimes \bm{v}_d]/2$. All calculations done, the vorticity at the centre of the sphere reflected by the Stokeslet is found to be zero (see the appendix \ref{sec:appendixC} for details). Only the flow due to the source-dipole creates a non-zero vorticity around the sphere. The torque-free condition requires the angular velocity of the sphere, $\bm{\Omega}$,  to be equal to the vorticity,
\begin{align}\label{eq:rotvel}
\bm{\Omega} = \bm{\omega} = -\frac{9 \mu_\alpha a^3}{16 \pi \mu_0 d_\alpha^7}    \hat{\bm{d}}_\alpha \btimes \bm{U} + \mathcal{O}(\mu^2_\alpha/\mu^2_0),
\end{align}
where $\bm{d}_\alpha = \bm{x}_\alpha - \bm{x}_c$, $d_\alpha = |\bm{d}_\alpha|$ and $\hat{\bm{d}}_\alpha= \bm{d}_\alpha/d_\alpha$. We assume there is an active mechanism that generates the translational velocity of the sphere such that it is attached to the body and rotates with it. The viscosity sink is assumed to be fixed in space, representing a viscosity gradient in space or produced by local heating of fluid. In the steady state, $\hat{\bm{d}}_\alpha$ and $\bm{U}$, become anti-parallel to each other. Hence, the active particle performs positive \textit{viscotaxis}, i.e.~it translates towards regions of higher viscosity \cite{petrino1978}.

We then introduce multiple viscosity sinks arranged in a hemispherical shell at $\bm{x} = \bm{d}_\alpha$, around one side of the sphere centred at the origin, so that $
\bm{d}_\alpha= d_\alpha [\sin\theta \cos\phi, \sin\theta \sin \phi, \cos\theta]$, where $\theta$ and $\phi$ are the polar and azimuthal angles, respectively. We can also define an effective viscosity gradient, $\bnabla \mu$, pointing along the negative $z-$ direction, see Fig.~\ref{fig2}a. The sphere's translational velocity, $\bm{U}$, makes an angle of $\Theta$ with $\bnabla \mu$ measured counter-clockwise. The individual contributions of the viscosity sinks are summed to find the net rotational velocity of the sphere,
\begin{align}
\Omega_y &=  - \beta \sin \Theta,
\end{align}
where
\begin{align}
\beta = \frac{9a^3\mu_\alpha U}{16\pi\mu_0 d_\alpha^7} \left[1 + \frac{N_\phi}{2} \left( \cot \frac{\pi}{4N_\theta} - 1\right) \right],
\end{align}
and $N_{\theta,\phi}$ are the total number of viscosity sinks along the polar and azimuthal directions such that $\theta \in [0,\pi/2)$ and $\phi \in [0,2\pi)$. 
The equation $\partial_t \Theta  = -\Omega_y$ is integrated in time to obtain,
\begin{align}
\Theta(t) = 2 \tan^{-1} \left[\tan \{ \Theta (0)/2) \} \exp{(\beta t)}\right].
\end{align}
The angle $\Theta(t)$ is plotted in Fig.~\ref{fig2}b for varying $d_\alpha^* = d_\alpha/a=1-2$ with 57 identical viscosities sinks of strength $\mu_\alpha = -0.01 \mu_0$ arranged around the particle in a hemisphere. It matches exactly with numerical solution obtained by integrating $\bm{U}$ in time as it rotates with $\bm{\Omega}$ (see the appendix \ref{sec:appendixC} for validation). For a given $N_\theta$, and area density of viscosity sinks, $n = [N_\phi(N_\theta-1) + 1]/4\pi d_\alpha^2$, we plot $\Omega_y$ as a function of $d_\alpha$ for a few different area densities in Fig.~\ref{fig2}c. Unsurprisingly, while $\Omega_y$ due to a single viscosity sink varies as $1/d_\alpha^{7}$ according to Eq.~\eqref{eq:rotvel}, integrating the contribution due to multiple sinks arranged in a hemispherical shell results in $\Omega_y \propto 1/d_\alpha^{5}$.

We compare these results with a recent article that has considered the effect of viscosity gradients, generated by temperature gradients, on the motion of rigid particles. \citet{oppenheimer2016} showed that the translational and rotational velocities of a Janus hot spherical particle gets coupled due to difference in viscosities around its surface. The hot side of the particle heats the fluid surrounding it and decreases its viscosity thereby creating a viscosity gradient.  Analytical progress was possible by introducing a small parameter, $\epsilon$, that captured small viscosity gradients and linearising the equations. The Faxen laws for a sphere in a fluid with a weakly varying linear viscosity gradient, $\mu(\bm{x}) = \mu_0 + \epsilon \bm{x} \bcdot \bnabla \mu$, were found to be, $\bm{F} = -6\pi\mu_0 a  \bm{U}  + \epsilon 2 \pi a^3 \bnabla  \mu \btimes \bm{\Omega}$ and $\bm{L} = -8\pi\mu_0 a^3  \bm{\Omega} - \epsilon 2 \pi a^3 \bnabla  \mu \btimes \bm{U}$. The angular velocity of a translating sphere is easily found from the torque-free condition by substituting $\bm{L} = \bm{0}$ to obtain, 
\begin{align}\label{eq:angveloppen}
\bm{\Omega} = - (\epsilon/ 4 \mu) \bnabla  \mu  \btimes \bm{U}.
\end{align}
Eq.~\eqref{eq:angveloppen} has the same functional form as Eq.~\eqref{eq:rotvel} for $\mu_\alpha < 0$ and $\bm{d}_\alpha = - \bnabla \mu$.  Consequently, it was shown that a torque-free translating sphere rotates such that its translational velocity vector will align with the viscosity gradient vector in the steady state which is consistent with our findings. Hence, we are able to obtain the same physics as \citet{oppenheimer2016} by modelling hot fluid around the sphere as concentrated viscosity sinks. Remarkably, the method of point viscosities can be used to create an arbitrary spatial viscosity distribution and applied to any arbitrary shaped particle.

The effect of a single point viscosity on the angular velocity of a translating sphere may appear weak due to its $1/d_\alpha^{7}$ dependency. This is because the disturbance vorticity has been calculated at the centre of the sphere using the far-field assumption which is valid when the point viscosity is far from the surface of the sphere. In order to explore the effect of a point viscosity accurately on the hydrodynamics of a rigid particle, while taking near-field effects into account, we perform simulations based on boundary element method \citep{pozrikidis1992,pozrikidis2002}. The point viscosity now interacts with the entire surface of the particle. The boundary integral equation relevant for Stokes equation with variable viscosity was derived by Pozrikidis using reciprocal theorem \cite{pozrikidis2016}. The volume integral involving variable viscosity was solved using finite element method in two-dimensions. For a point viscosity in space, the equations simplify considerably (see the appendix \ref{sec:appendixD} for derivation). The hydrodynamics of a rigid particle in the presence of point viscosities is written succinctly in the form of an integral equation, 
\begin{align}\label{eq:bie}
\bm{v}(\bm{x}_0) = -\frac{1}{8\pi\mu_0} \iint_S \bm{f}(\bm{x},\bm{x}_\alpha) \bcdot \mathsfbi{G}_{mod}(\bm{x},\bm{x}_0,\bm{x}_\alpha) \, \mathrm{d}S(\bm{x}), 
\end{align}
where $\bm{x}_0\neq \bm{x}_\alpha$ is an evaluation point anywhere in the fluid domain including the sphere's surface, $\bm{x}\in S$ is integration point on the sphere's surface, $\mathsfbi{G}_{mod}= \mathsfbi{G} - \mathsfbi{G}_v$ is the modified Green's function, $\mathsfbi{G}$ is the free-space Green's function, $G_{v,lj} = (\mu_\alpha/8\pi\mu_0) T_{ijk}(\bm{x}_\alpha,\bm{x}_0)  [\bnabla_{y,i} G_{lk}(\bm{x},\bm{y})]_{\bm{y}=\bm{x}_\alpha}$ is the Green's function due to a point viscosity, and $\mathsfbi{T}(\bm{x}_\alpha,\bm{x}_0)  = -6 \bm{p}\bm{p}\bm{p}/p^5$ is the stresslet, where $\bm{p} = \bm{x}_\alpha - \bm{x}_0$. The surface velocity $\bm{v}(\bm{x}) = \bm{U} + \bm{\Omega} \btimes (\bm{x} - \bm{x}_c)$ is prescribed by rigid body motion and $\bm{f}$ is the hydrodynamic traction acting on the body. 
\begin{figure}
    \centering
    \includegraphics[width=0.9\linewidth]{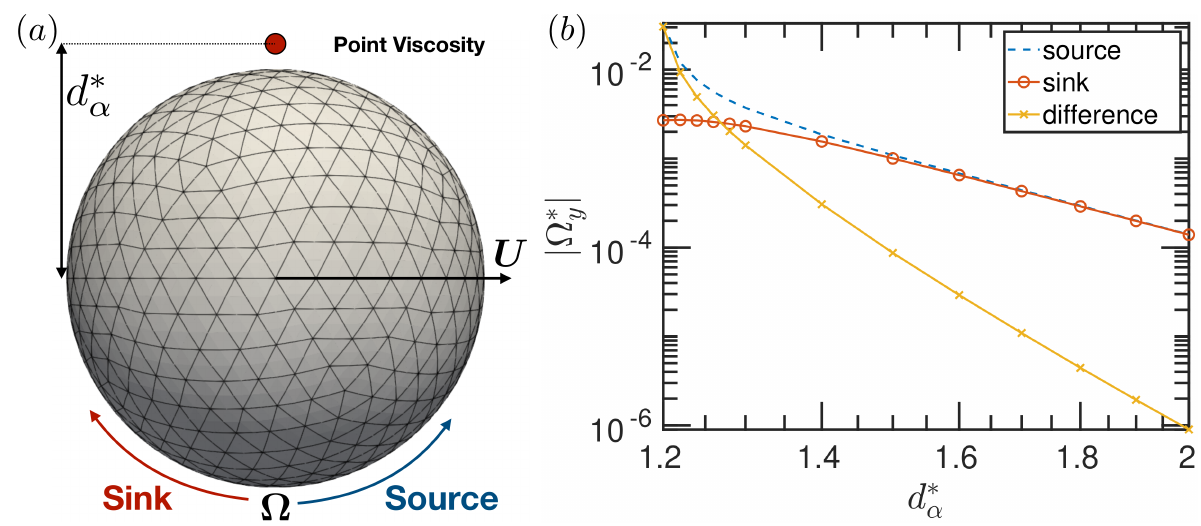}
    \caption{Log-log plot of the absolute value of the angular velocity, $|\Omega_y^*|$, of a translating spherical particle in the presence of a viscosity source (blue-dashed) and sink (red-circle) of strength $\mu_\alpha = \pm 0.1\mu_0$, solved using boundary element method as a function of the distance between the sphere centre and the point viscosity, denoted as $d_\alpha^*$. The difference between the magnitude of the angular velocities due to the source and sink (yellow-cross) is seen to decrease as they move farther away from the sphere's surface.}
    \label{fig3}
\end{figure}

We again consider the hydrodynamics of a torque-free sphere, where we impose a translational velocity and find its rotational velocity due to the presence of a point viscosity, see Fig.~\ref{fig3}(a). The viscous force and torque acting on the sphere are found by integrating the hydrodynamic tractions after solving Eq.\eqref{eq:bie} numerically by discretising the sphere's surface into triangular elements \citep{pozrikidis1992,pozrikidis2002}, $
\bm{F}  = \iint_S  \bm{f}\, \mathrm{d}S(\bm{x}),  ~ \bm{L}  = \iint_S (\bm{x} - \bm{x}_c)\btimes \bm{f}\, \mathrm{d}S(\bm{x})$.
The sphere is centred at the origin, $\bm{x}_c=\bm{0}$ and the point visoscity is placed at a distance $\bm{d}^* = (0,0,d_\alpha^*)$, scaled by the sphere radius, $a$. The magnitude of the angular velocity of the sphere, $\bm{\Omega} = -\bm{L}/8\pi\mu_0 a^3$, is plotted for the case of a single isolated viscosity sink and source in Fig.~\ref{fig3}(b).  Here, we make a curious observation. The angular velocity due to a viscosity sink and source are found to be in the positive and negative $y-$  direction, respectively, as expected. However, their magnitudes are the not the same when $d_\alpha^* <1.5$. This is in contrast to both the results of \citet{oppenheimer2016}, see Eqs.~\eqref{eq:angveloppen}, and the far-field result derived in this paper, see Eq.~\eqref{eq:rotvel}, wherein switching the direction of viscosity gradient or changing a sink into a source simply changes the direction of the angular velocity, while its magnitude remains unchanged.  This difference arises because both Eqs.~\eqref{eq:angveloppen} and Eq.~\eqref{eq:rotvel} only consider the leading order viscosity gradient and point viscosity effects, respectively. In contrast, in the numerical simulations, the point viscosity interacts nonlinearly with the sphere's surface. As a result, the magnitude of the angular velocity approach each other only when $||\mathsfbi{G}_v||$ becomes sufficiently small compared to $||\mathsfbi{G}||$, i.e. when the point viscosity is far away from the surface or is sufficiently weak in magnitude. This is easily seen by considering two cases: hydrodynamics of a sphere placed next to a viscosity source (subscript 1) and a sink (subscript 2). The resulting system of linear equations arising from Eq.~\eqref{eq:bie} for these two cases are $ \bm{f}_1 = (\mathsfbi{G} - \mathsfbi{G}_v)^{-1} \bm{v}$,  $ \bm{f}_2 = (\mathsfbi{G} + \mathsfbi{G}_v)^{-1} \bm{v}$, where the vectors $\bm{v}, \bm{f}$ denote the surface velocity and traction values on the discretised sphere and $\mathsfbi{G}$ denotes the Green's function matrix.  Integrating the tractions gives us the hydrodynamic torque, $\bm{L}_{1,2}  = \iint_S (\bm{x} - \bm{x}_c)\btimes [(\mathsfbi{G} \pm \mathsfbi{G}_v)^{-1} \bm{v}]\, \mathrm{d}S(\bm{x})$. Noting that $\iint_S (\bm{x} - \bm{x}_c)\btimes \mathsfbi{G}^{-1} \bm{v} \, \mathrm{d}S(\bm{x}) = \bm{0}$, and doing a Taylor series expansion of the torque, we find, $ \bm{L}_1 \rightarrow - \bm{L}_2$,  as $\mathsfbi{G}^{-1} \mathsfbi{G}_v$ becomes smaller, thereby, explaining the conundrum.

In this Letter, we proposed a novel method to model viscosity heterogeneities  in a fluid shrunk to a point. Specifically, we asked: how are the velocity and pressure fields affected by the presence of a point viscosity variation subject to a background flow? We found the disturbance flow field to be the same as that due to a stresslet, written as a singularity solution. It is worth noting that a wide variety of physical problems like potential flow, electrostatics, linear elasticity, wave propagation, and viscous flow  are amenable to theoretical analysis because their governing equations admit singularity solutions. Hence, the ideas presented here may be applied to these other physical phenomena as well when material heterogeneities are present in the media.

The modelling framework developed in this Letter opens up several new avenues of research. One such avenue is modelling generalized non-Newtonian fluids relevant for problems in biological fluids. For example, it has been hypothesised that the gastric pathogen, \textit{Helicobacter pylori}, is able to propel itself through the mucus gel by reducing its viscoelasticity and attach to epithelial cells \citep{celli2009}. Simplified theoretical models to understand this phenomena have been developed based on Taylor's swimming sheet in a phase-separated fluid \citep{man2015} and in a layer of Newtonian fluid bounded by a Brinkman fluid \cite{mirbagheri2016}. Using the point viscosity model, solving the hydrodynamics of a fully three-dimensional model of a bacterium swimming \citep{das2018,das2019} in a heterogeneous fluid becomes feasible. Recent articles have considered the effect of viscosity gradients on swimming microorganisms like green algae \citep{stehnach2021,coppola2021} and model active swimmers \citep{liebchen2018,datt2019}. Incorporating the effect of point viscosities in such model swimmers will yield further physical insight into how microorganisms respond to viscosity variations in the fluid environments where they live. Also, the effect of nonlinear interactions between multiple point viscosities on a particle's motion has been neglected here, and may have nontrivial consequences. Finally, the point viscosities may be made to diffuse in time and advect with the velocity field around them, thereby relaxing the quasi-static assumption. 


\appendix
\onecolumngrid

\section{Detailed derivation of the flow velocity and pressure fields due to point viscosity variations}\label{sec:appendixA}

Consider Stokes equation with a variable viscosity field, $\mu (\bm{x}) = \mu_0 + \sum_{\alpha=1}^N \mu_\alpha \delta (\bm{x}-\bm{x}_\alpha)$,
\begin{align}
&-\bnabla p(\bm{x}) + \bnabla \bcdot [\mu (\bm{x}) \{ \bnabla \bm{v}(\bm{x}) + \bnabla \bm{v}^T(\bm{x}) \} ] +  \bm{F} \delta(\bm{x} - \bm{x}_0) = \bm{0}, \quad \bnabla \bcdot \bm{v} = 0.
\end{align}
We solve this equation for any point in space for which $\bm{x} \neq \bm{x}_0, \bm{x} \neq \bm{x}_\alpha$, $\alpha \in [1,N]$. If $\bm{x} = \bm{x}_\alpha$, i.e. we wish to find the flow velocity and pressure at the location of a point viscosity, we do not consider that point viscosity in the governing equation, as it cannot induce a flow at its own location. The governing equation then takes the form,
\begin{align}
\begin{split}
&-\bnabla p(\bm{x}) +  \sum_{\alpha=1}^N \mu_\alpha  \bnabla \delta (\bm{x}-\bm{x}_\alpha) \bcdot [ \bnabla \bm{v}(\bm{x}) + \bnabla \bm{v}^T(\bm{x}) ] \\
& + [\mu_0 + \sum_{\alpha=1}^N \mu_\alpha \delta  (\bm{x}-\bm{x}_\alpha)] \nabla^2  \bm{v}(\bm{x}) +  \bm{F} \delta(\bm{x} - \bm{x}_0) = \bm{0},
\end{split}
\end{align}
We wish to use Fourier Transform to solve the above equation,
\begin{align}
\begin{split}
& (p,\bm{v}) = (2\pi)^{-3/2} \displaystyle\int_{-\infty}^\infty (\hat{p},\hat{\bm{v}})   ~ \exp{ (i\bm{k}\bcdot\bm{x})} ~\mathrm{d}\bm{k}, \\
& (\hat{p},\hat{\bm{v}})  = (2\pi)^{-3/2} \displaystyle\int_{-\infty}^\infty (p,\bm{v})  ~ \exp{ (-i\bm{k}\bcdot\bm{x})} ~\mathrm{d}\bm{x}.
\end{split}
\end{align}
Let us first find the Fourier Transform of  $ \sum_{\alpha=1}^N \mu_\alpha  \bnabla \delta (\bm{x}-\bm{x}_\alpha) \bcdot [\bnabla \bm{v}(\bm{x}) +\bnabla \bm{v}^T(\bm{x})]$, denoting it is as $\hat{a}$,
\begin{align}
\begin{split}
\hat{a} &=  \frac{1}{(2\pi)^{3/2}} \sum_{\alpha=1}^N  \int_{-\infty}^\infty \mu_\alpha  \bnabla \delta (\bm{x}-\bm{x}_\alpha) \bcdot [ \bnabla \bm{v}(\bm{x}) + \bnabla \bm{v}^T(\bm{x}) ]  ~ \mathrm{e}^{ -i\bm{k}\bcdot\bm{x} } ~\mathrm{d}\bm{x}, \\
&= - \frac{1}{(2\pi)^{3/2}} \sum_{\alpha=1}^N  \mu_\alpha  [ \nabla^2 \bm{v}(\bm{x})    - i \bm{k} \bcdot \{ \bnabla \bm{v}(\bm{x}) + \bnabla \bm{v}^T(\bm{x}) \} ]_{\bm{x}=\bm{x}_\alpha} ~\mathrm{e}^{ -i\bm{k}\bcdot\bm{x}_\alpha }
\end{split}
\end{align}
Similarly, let us find the Fourier Transform of the next term $[\mu_0 + \sum_{\alpha=1}^N \mu_\alpha \delta  (\bm{x}-\bm{x}_\alpha)] \nabla^2  \bm{v}(\bm{x})$,  denoting it is as $\hat{b}$,
\begin{align}
\begin{split}
\hat{b} &=  \frac{1}{(2\pi)^{3/2}}  \int_{-\infty}^\infty  \left[\mu_0 + \sum_{\alpha=1}^N\mu_\alpha \delta  (\bm{x}-\bm{x}_\alpha)\right] \nabla^2   \bm{v}(\bm{x}) ~ \mathrm{e}^{ -i\bm{k}\bcdot\bm{x} } ~\mathrm{d}\bm{x},  \\
&=  \frac{1}{(2\pi)^{3/2}} \int_{-\infty}^\infty  \mu_0 \nabla^2   \bm{v}(\bm{x}) ~ \mathrm{e}^{ -i\bm{k}\bcdot\bm{x} } ~\mathrm{d}\bm{x} + \frac{1}{(2\pi)^{3/2}} \sum_{\alpha=1}^N \mu_\alpha  [\nabla^2   \bm{v}(\bm{x})]_{\bm{x}=\bm{x}_\alpha}~\mathrm{e}^{ -i\bm{k}\bcdot\bm{x}_\alpha } \\
&= - \mu_0 k^2  \hat{\bm{v}}  + \frac{1}{(2\pi)^{3/2}} \sum_{\alpha=1}^N \mu_\alpha [\nabla^2   \bm{v}(\bm{x})]_{\bm{x}=\bm{x}_\alpha}~\mathrm{e}^{ -i\bm{k}\bcdot\bm{x}_\alpha }
\end{split}
\end{align}
Finally,  we need to find the Fourier Transform of the point force,
\begin{align}
\begin{split}
\hat{\bm{F}}  = \frac{1}{(2\pi)^{3/2}}\int_{-\infty}^\infty \bm{F} \delta (\bm{x}  - \bm{x}_0) ~ \mathrm{e}^{- i\bm{k}\bcdot\bm{x} } ~\mathrm{d}\bm{x} =  \frac{1}{(2\pi)^{3/2}} \bm{F}  ~ \mathrm{e}^{- i\bm{k}\bcdot \bm{x}_0 } 
\end{split}
\end{align}
Putting it all together, and noting that one term in $\hat{a}$ and $\hat{b}$ cancel each other, the Fourier Transform of the Stokes equation is,
\begin{align}
\begin{split}
 \frac{1}{(2\pi)^{3/2}}\int_{-\infty}^\infty &\left[- i \bm{k} \hat{p} -  \mu_0 k^2  \hat{\bm{v}} +  \frac{1}{(2\pi)^{3/2}} \sum_{\alpha=1}^N \mu_\alpha \mathrm{e}^{ -i\bm{k}\bcdot\bm{x}_\alpha } i \bm{k} \bcdot \{ \bnabla \tilde{\bm{v}}(\bm{x}) + \bnabla \tilde{\bm{v}}^T(\bm{x}) \} \Big|_{\bm{x}= \bm{x}_\alpha} \right. \\
&  \left.  + \hat{\bm{F}}  \right]~ \mathrm{e}^{ i\bm{k}\bcdot\bm{x}} ~\mathrm{d}\bm{k} = \bm{0}, 
\end{split}
\end{align}
together with $ i \bm{k} \bcdot \hat{\bm{v}}= 0$.  Note that the velocity to be evaluated at $\bm{x} = \bm{x}_\alpha$ is independent of $\bm{k}$ and includes the velocity contribution from the point force and all point viscosities except the one at $\bm{x} = \bm{x}_\alpha$.  To make this distinction, we have replaced $\bm{v}$ with $\tilde{\bm{v}}$ and denote $ \{ \bnabla \tilde{\bm{v}}(\bm{x}) + \bnabla \tilde{\bm{v}}^T(\bm{x}) \} \Big|_{\bm{x}= \bm{x}_\alpha}  = 2\bm{E}_\alpha$ which is simply the straining flow at the location of the point viscosity. 
\begin{align}
- i \bm{k} \hat{p} +  i  \sum_{\alpha=1}^N \frac{ \mu_\alpha }{(2\pi)^{3/2}}  \bm{k} \bcdot 2\bm{E}_\alpha \mathrm{e}^{ -i\bm{k}\bcdot\bm{x}_\alpha } - \mu_0 k^2  \hat{\bm{v}}  + \hat{\bm{F}}  = \bm{0}
\end{align}
Taking a dot product with $\bm{k}$ and using $\bm{k} \bcdot \hat{\bm{v}}= 0$, we get an expression for $\hat{p}$,
\begin{align}
\hat{p}   =  \sum_{\alpha=1}^N \frac{ \mu_\alpha }{(2\pi)^{3/2}} \frac{(\bm{k} \bcdot 2\bm{E}_\alpha \bcdot \bm{k}) \mathrm{e}^{ -i\bm{k}\bcdot\bm{x}_\alpha} }{ k^2} - i \frac{\bm{k} \bcdot \hat{\bm{F}}}{k^2} = \sum_{\alpha=1}^N \frac{\mu_\alpha}{(2\pi)^{3/2}} 2 \hat{P}_{ij} E_{\alpha,ij} \mathrm{e}^{ -i\bm{k}\bcdot\bm{x}_\alpha } - i \hat{Q}_i \hat{F}_i,
\end{align}
where $\hat{P}_{ij} = k_i k_j/k^2$ and $\hat{Q}_{i} = k_i/k^2$. The solution to the pressure field is found by taking the inverse Fourier Transform,
\begin{align}
p &=  \sum_{\alpha=1}^N \frac{\mu_\alpha}{(2\pi)^3} 2E_{\alpha,ij} \int_{-\infty}^\infty \hat{P}_{ij} ~ \mathrm{e}^{ i\bm{k}\bcdot (\bm{x}-\bm{x}_\alpha)}  ~\mathrm{d}\bm{k} - \frac{1}{(2\pi)^3} F_i \int_{-\infty}^\infty  i \hat{Q}_{i} ~ \mathrm{e}^{ i\bm{k}\bcdot(\bm{x} - \bm{x}_0) }  ~\mathrm{d}\bm{k}.
\end{align}
We know from the fundamental solution of Laplace's equation \cite{kim2013,lisicki2013},
\begin{align}
{\frac{1}{4\pi r_\alpha}} &= \frac{1}{(2\pi)^3} \int_{-\infty}^\infty \frac{1}{k^2} ~ \mathrm{e}^{ i\bm{k}\bcdot\bm{r}_\alpha } ~\mathrm{d}\bm{k},
\end{align}
where $\bm{r}_\alpha = \bm{x} - \bm{x}_\alpha$, and $\bm{r}_0 = \bm{x} - \bm{x}_0$. Taking gradients of both sides w.r.t $\bm{x}$, we get,
\begin{align}
\bnabla \left( \frac{1}{4\pi r_\alpha} \right) &= \frac{1}{(2\pi)^3} \int_{-\infty}^\infty i \frac{ \bm{k} }{k^2} ~ \mathrm{e}^{ i\bm{k}\bcdot\bm{r}_\alpha } ~\mathrm{d}\bm{k}, \quad \bnabla \bnabla \left( \frac{1}{4\pi r_\alpha} \right) &= -\frac{1}{(2\pi)^3} \int_{-\infty}^\infty \frac{\bm{k}\bm{k}}{k^2} ~ \mathrm{e}^{ i\bm{k}\bcdot\bm{r}_\alpha } ~\mathrm{d}\bm{k}.
\end{align}
This gives us the desired pressure field,
\begin{align}
\begin{split}
p  &= - \bnabla \left( \frac{1}{4\pi r_0} \right) \bcdot \bm{F} -\sum_{\alpha=1}^N \mu_\alpha \bnabla \bnabla \left( \frac{1}{4\pi r_\alpha} \right) \bm{:} 2\bm{E}_\alpha, \\
& = \frac{\bm{r}_0 \bcdot \bm{F}}{4 \pi r_0^3} - \sum_{\alpha=1}^N \frac{3 \mu_\alpha}{2\pi r_\alpha^5} [\bm{r}_\alpha\bm{r}_\alpha \bm{:} (\bnabla \tilde{\bm{v}})_{\bm{x} = \bm{x}_\alpha}]
\end{split}
\end{align}
Substituting the expression for $\hat{p}$ in the momentum balance equation,
\begin{align}
\begin{split}
\hat{v}_i &=  \frac{1}{\mu_0 k^2}\left[ \sum_{\alpha=1}^N \frac{i \mu_\alpha}{(2\pi)^{3/2}} \left( \delta_{im} - \frac{k_i k_m}{k^2}\right) 2 k_j E_{\alpha,jm}  \mathrm{e}^{ -i\bm{k}\bcdot\bm{x}_\alpha} + \left( \delta_{im} -  \frac{k_i k_m}{k^2}\right)   \frac{1}{(2\pi)^{3/2}}  F_m ~ \mathrm{e}^{- i\bm{k}\bcdot \bm{x}_0  }  \right] \\
& = \frac{1}{(2\pi)^{3/2}}  \left(\sum_{\alpha=1}^N  \frac{\mu_\alpha}{\mu_0} 2\hat{\mathcal{G}}_{\alpha, imj} E_{\alpha,jm}  \mathrm{e}^{ -i\bm{k}\bcdot\bm{x}_\alpha} +   \frac{1}{\mu_0}  \hat{\mathcal{H}}_{im}  F_m  \mathrm{e}^{- i\bm{k}\bcdot \bm{x}_0 } \right),
\end{split}
\end{align}
where,
\begin{align}
\hat{\mathcal{G}}_{\alpha, imj} = \left( \delta_{im} - \frac{k_i k_m}{ k^2 }  \right) \frac{i k_j}{k^2 }, \qquad \hat{\mathcal{H}}_{ im} = \left( \delta_{im} -  \frac{k_i k_m}{ k^2 }  \right) \frac{1}{k^2 }.
\end{align}
Now, let us take the inverse Fourier Transform of the above equation,
\begin{align}
\bm{v}= \frac{1}{(2\pi)^3} \left[ \sum_{\alpha=1}^N \frac{\mu_\alpha}{ \mu_0} 2E_{\alpha,jm} \int_{-\infty}^\infty \hat{\mathcal{G}}_{\alpha, imj}  ~ \mathrm{e}^{ i\bm{k}\bcdot (\bm{x} -\bm{x}_\alpha)} ~\mathrm{d}\bm{k} + \frac{1}{ \mu_0} F_m \int_{-\infty}^\infty \hat{\mathcal{H}}_{im}  ~ \mathrm{e}^{ i\bm{k}\bcdot(\bm{x} -\bm{x}_0)} ~\mathrm{d}\bm{k}  \right].
\end{align}
We know from the fundamental solution to Stokes equation,
\begin{align}
\frac{1}{8\pi} \left(\frac{\mathcal{I}}{r} + \frac{\bm{r} \bm{r} }{r^3} \right)&= \frac{1}{(2\pi)^3}\int_{-\infty}^\infty \left(  \frac{\mathcal{I}}{k^2} - \frac{\bm{k}\bm{k}}{ k^4}  \right)~ \mathrm{e}^{ i\bm{k}\bcdot\bm{x} } ~\mathrm{d}\bm{k}.
\end{align}
We take the gradient of the above equation,
\begin{align}
\frac{1}{8\pi} \bnabla \left(\frac{\mathcal{I}}{r} + \frac{\bm{r} \bm{r} }{r^3} \right)&= \frac{1}{(2\pi)^3}\int_{-\infty}^\infty  i \bm{k} \left(  \frac{\mathcal{I}}{k^2} - \frac{\bm{k}\bm{k}}{ k^4}  \right)~ \mathrm{e}^{ i\bm{k}\bcdot\bm{x} } ~\mathrm{d}\bm{k}.
\end{align}
The velocity field in the real space is then,
\begin{align}
v_i = \sum_{\alpha=1}^N \frac{\mu_\alpha}{8\pi\mu_0} \frac{\partial }{\partial x_j} \left(\frac{\delta_{im}}{r_\alpha} + \frac{r_{\alpha,i}r_{\alpha,m}}{r_\alpha^3} \right) (2E_{\alpha,jm}) + \frac{1}{8\pi\mu_0}  \left(\frac{\delta_{im}}{r_0} + \frac{r_{0,i}r_{0,m}}{r_0^3} \right) F_m
\end{align}
where $2\bm{E}_\alpha =  \{ \bnabla \bm{v}(\bm{x}) + \bnabla \bm{v}^T(\bm{x}) \}_{\bm{x}=\bm{x}_\alpha} $ which can be simplified to get the desired expression for the velocity field due to a point force and point viscosities,
\begin{align}
\bm{v} = \frac{1}{8\pi\mu_0}  \left(\frac{\bm{I}}{r_0} + \frac{\bm{r}_0\bm{r}_0 }{r_0^3} \right) \bcdot \bm{F} - \sum_{\alpha=1}^N \frac{3\mu_\alpha}{4\pi\mu_0 r_\alpha^5}  \bm{r}_\alpha[\bm{r}_\alpha\bm{r}_\alpha \bm{:} \bnabla \tilde{\bm{v}}|_{\bm{x} = \bm{x}_\alpha}] 
\end{align}

\section{Interactions between point viscosities}\label{sec:appendixB}

\subsection{Method of reflection}
Let us assume, we have three point viscosities in space, $\alpha \in [1,2,3]$ and the background velocity be denoted as $\bm{v}_0$. We first consider the interaction of the background velocity with the point viscosity, neglecting any interactions between them. The leading order velocity disturbance at any point due to the presence of the point viscosities considered individually are,
\begin{subequations}\label{eq:zerothorder}
\begin{align}
\bm{v}_{1}^{0} &= - \frac{3\mu_1}{4\pi\mu_0 r_1^5}  \bm{r}_1 \bm{r}_1\bm{r}_1 \bm{:} \{ \bnabla \bm{v}_0\}_{\bm{x} = \bm{x}_1}, \\
\bm{v}_{2}^{0} &=  - \frac{3\mu_2}{4\pi\mu_0 r_2^5}  \bm{r}_2 \bm{r}_2\bm{r}_2 \bm{:} \{ \bnabla \bm{v}_0\}_{\bm{x} = \bm{x}_2}, \\
\bm{v}_{3}^{0} &= - \frac{3\mu_3}{4\pi\mu_0 r_3^5}  \bm{r}_3 \bm{r}_3\bm{r}_3 \bm{:} \{ \bnabla \bm{v}_0\}_{\bm{x} = \bm{x}_3},
\end{align}
\end{subequations}
where the subscript denotes which point viscosity variation is disturbing the background flow and the superscript denotes the asymptotic order. Having found the leading order velocities, we can correct the velocities, accounting for interactions between the point viscosities,
\begin{subequations}\label{eq:firstorder}
\begin{align}
\bm{v}_{1}^{1} &= - \frac{3\mu_1}{4\pi\mu_0 r_1^5}  \bm{r}_1 \bm{r}_1\bm{r}_1 \bm{:} \{ \bnabla ( \bm{v}_2^{0} + \bm{v}_3^{0}) \}_{\bm{x} = \bm{x}_1}, \\
\bm{v}_{2}^{1} &= - \frac{3\mu_2}{4\pi\mu_0 r_2^5}  \bm{r}_2 \bm{r}_2\bm{r}_2 \bm{:} \{ \bnabla ( \bm{v}_3^{0} + \bm{v}_1^{0}) \}_{\bm{x} = \bm{x}_2}, \\
\bm{v}_{3}^{1} &= - \frac{3\mu_3}{4\pi\mu_0 r_3^5}  \bm{r}_3 \bm{r}_3\bm{r}_3 \bm{:} \{ \bnabla (\bm{v}_1^{0} + \bm{v}_2^{0}) \}_{\bm{x} = \bm{x}_3}.
\end{align}
\end{subequations}
The flow velocity at any point when $\bm{x} \neq \bm{x}_{1,2,3}$ is then given as,
\begin{align}
\bm{v} = \bm{v}_0 + \bm{v}_{1}^{0} + \bm{v}_{1}^{1} + \bm{v}_{1}^{2} + \ldots + \bm{v}_{2}^{0} + \bm{v}_{2}^{1} + \bm{v}_{2}^{2} + \ldots + \bm{v}_{3}^{0} + \bm{v}_{3}^{1} + \bm{v}_{3}^{2} + \ldots.
\end{align}
The flow velocity at the location of a point viscosity, say $\bm{x} = \bm{x}_{1}$ is,
\begin{align}
\bm{v} = \bm{v}_0 + \bm{v}_{2}^{0} + \bm{v}_{2}^{1} + \bm{v}_{2}^{2} + \ldots + \bm{v}_{3}^{0} + \bm{v}_{3}^{1} + \bm{v}_{3}^{2} + \ldots,
\end{align}
where hydrodynamic interactions occur between point viscosities 2 and 3 only,
\begin{subequations}\label{eq:only2and3}
\begin{align}
&\bm{v}_{2}^{0} + \bm{v}_{2}^{1} = - \frac{3\mu_2}{4\pi\mu_0 r_2^5}  \bm{r}_2 \bm{r}_2\bm{r}_2 \bm{:} \{ \bnabla (\bm{v}_0 + \bm{v}_3^{0} ) \}_{\bm{x} = \bm{x}_2}, \\
&\ \bm{v}_{3}^{0}  + \bm{v}_{3}^{1} = - \frac{3\mu_3}{4\pi\mu_0 r_3^5}  \bm{r}_3 \bm{r}_3\bm{r}_3 \bm{:} \{ \bnabla (\bm{v}_0 + \bm{v}_2^{0}) \}_{\bm{x} = \bm{x}_3}.
\end{align}
\end{subequations}
Simply substituting $\mu_1 =0$ in Eq.~\eqref{eq:firstorder}, is an easy way to reproduce Eq.~\eqref{eq:only2and3}. Similar equations can be obtained when we wish to compute the velocity field at the location of other point viscosities $\bm{x} = \bm{x}_{2,3}$.

\subsection{Exact method}
Let us assume, we have three point viscosities in space, $\alpha \in [1,2,3]$ and the background velocity be denoted as $\bm{v}_0$. The velocity field is,
\begin{align}\label{eq:veloriginal}
\bm{v} =  \bm{v}_0 - \sum_{\alpha=1}^N \frac{3\mu_\alpha}{4\pi\mu_0 r_\alpha^5}  \bm{r}_\alpha[\bm{r}_\alpha\bm{r}_\alpha \bm{:} \bnabla \tilde{\bm{v}}|_{\bm{x} = \bm{x}_\alpha}] 
\end{align}
Taking gradient of the equation above, we get,
\begin{align}
\bnabla \bm{v} =  \bnabla \bm{v}_0 + \sum_{\alpha=1}^N  \frac{\mu_\alpha}{8\pi\mu_0 }  \bnabla \left[ -6 \frac{\bm{r}_\alpha \bm{r}_\alpha \bm{r}_\alpha}{r_\alpha^5} \right] \bm{:} \bnabla \tilde{\bm{v}}|_{\bm{x} = \bm{x}_\alpha}.
\end{align}
Denoting the gradient of the stresslet as a fourth-order tensor, $\mathsfbi{S}= \bnabla [-6 \bm{r}_\alpha \bm{r}_\alpha \bm{r}_\alpha/r_\alpha^5]$, we find the gradients at the location of the point viscosities to obtain coupled nonlinear equations,
\begin{subequations}\label{eq:grads}
\begin{align}
& \bnabla \tilde{\bm{v}}|_{\bm{x} = \bm{x}_1}  - \frac{\mu_2}{8\pi\mu_0} \mathsfbi{S}_2(\bm{x}_1) \bm{:} \bnabla \tilde{\bm{v}}|_{\bm{x} = \bm{x}_2} - \frac{\mu_3}{8\pi\mu_0} \mathsfbi{S}_3(\bm{x}_1) \bm{:} \bnabla \tilde{\bm{v}}|_{\bm{x} = \bm{x}_3} =  \bnabla \bm{v}_0(\bm{x}_1), \\
- &\frac{\mu_2}{8\pi\mu_0} \mathsfbi{S}_1(\bm{x}_2) \bm{:} \bnabla \tilde{\bm{v}}|_{\bm{x} = \bm{x}_1} + \bnabla \tilde{\bm{v}}|_{\bm{x} = \bm{x}_2} - \frac{\mu_3}{8\pi\mu_0} \mathsfbi{S}_3(\bm{x}_2) \bm{:} \bnabla \tilde{\bm{v}}|_{\bm{x} = \bm{x}_3} =  \bnabla \bm{v}_0(\bm{x}_2), \\
- &\frac{\mu_2}{8\pi\mu_0} \mathsfbi{S}_1(\bm{x}_3) \bm{:} \bnabla \tilde{\bm{v}}|_{\bm{x} = \bm{x}_1} - \frac{\mu_2}{8\pi\mu_0} \mathsfbi{S}_2(\bm{x}_3) \bm{:} \bnabla \tilde{\bm{v}}|_{\bm{x} = \bm{x}_2} + \bnabla \tilde{\bm{v}}|_{\bm{x} = \bm{x}_3} =  \bnabla \bm{v}_0(\bm{x}_3).
\end{align}
\end{subequations}
that can be solved for the velocity gradients at the location of the point viscosities, $\bnabla \tilde{\bm{v}}|_{\bm{x} = \bm{x}_\alpha}$. We can substitute them back in Eq.~\eqref{eq:veloriginal} to find the velocity at any point in space.

\section{Translating sphere in the presence of point viscosities}\label{sec:appendixC}

The flow field due to a translating sphere in the absence of any point viscosities is,
\begin{align}
v_{0,i} = \frac{3a}{4} U_j\left(\frac{\delta_{ij}}{r_c} + \frac{r_{i,c} r_{j,c}}{r_c^3} \right) +\frac{3a^3}{4} U_j\left(\frac{\delta_{ij}}{3r_c^3} - \frac{r_{i,c} r_{j,c}}{r_c^5} \right),
\end{align}
where $\bm{r}_c = \bm{x} - \bm{x}_c$. Let us first consider the Stokeslet term and find its velocity gradient,
\begin{align}
\frac{\partial v_{0,i}}{\partial x_k} = \frac{3a}{4} U_j\frac{\partial}{\partial x_k}\left(\frac{\delta_{ij}}{r_c} + \frac{r_{i,c} r_{j,c}}{r_c^3} \right) = \frac{3a}{4} U_j \left(\frac{-\delta_{ij}r_{k,c} + \delta_{ik} r_{j,c} + r_{i,c} \delta_{jk}}{r_c^3} - \frac{3r_{i,c} r_{j,c} r_{k,c}}{r_c^5} \right).
\end{align}
The disturbance velocity due to a single isolated point viscosity is,
\begin{subequations}
\begin{align}
 \bm{v} &= -\frac{3\mu_1}{4\pi\mu_0 r_1^5}  \bm{r}_1\bm{r}_1\bm{r}_1 \bm{:} (\bnabla \bm{v}_0)_{\bm{x} =   \bm{x}_1},\\
v_m &= - \frac{9a U_j\mu_1}{32\pi\mu_0 } U_j\frac{r_{1,m}  r_{1,i} r_{1,k}}{r_1^5} \left[ \frac{-\delta_{ij}d_k  + \delta_{ik}d_j + \delta_{jk}d_i }{d^3} -\frac{3 d_i d_j d_k}{d^5}\right].
\end{align}
\end{subequations}
Above, we substituted $\bm{x} = \bm{x}_1$ in $\bm{r}_c = \bm{x} - \bm{x}_c$ and denote $\bm{d} = \bm{x}_1 - \bm{x}_c$ as the vector pointing from the sphere centre to the point viscosity.
The disturbance vorticity induced at the particles location is,
\begin{align}
\begin{split}
\omega_{p} & =  - \frac{9a U_j\mu_1}{32\pi\mu_0 } U_j\frac{1}{2} \left[ \frac{r_{1,m}  \epsilon_{pim} r_{1,k}}{r_1^5} + \frac{r_{1,m}  r_{1,i} \epsilon_{pkm}}{r_1^5} \right] \left[ \frac{-\delta_{ij}d_k  + \delta_{ik}d_j + \delta_{jk}d_i }{d^3} -\frac{3 d_i d_j d_k}{d^5}\right].
\end{split}
\end{align}
Substituting $\bm{x} = \bm{x}_c$ in $\bm{r}_1 = \bm{x}_c - \bm{x}_1  = -\bm{d}$, we find $\bm{\omega} = \bm{0}$. Next, we consider the source-dipole term and find the velocity gradient,
\begin{align}
\frac{\partial v_{0,i}}{\partial x_k} = \frac{a^3}{4} U_j\frac{\partial}{\partial x_k}\left(\frac{\delta_{ij}}{r_c^3} - 3\frac{r_{i,c} r_{j,c}}{r_c^5} \right) =  -\frac{3a^3}{4} U_j \left(\frac{\delta_{ij}r_{k,c}}{r_c^5} + \frac{\delta_{ik} r_{j,c}}{r_c^5} + \frac{r_{i,c}\delta_{jk}}{r_c^5} -5 \frac{r_{i,c} r_{j,c} r_{k,c}}{r_c^7} \right).
\end{align}
For convenience, let us denote $\partial v_{0,i}/\partial x_k = -(3a^3/4) S_{ik}$. The disturbance velocity due to the point viscosity is,
\begin{align}
v_m &=  \left[ -\frac{3\mu_1}{4\pi\mu_0 r_1^5}  r_{1,m} r_{1,i} r_{1,k} \right]  \left[-\frac{3a^3}{4} S_{ik} \right] =  \frac{9a^3\mu_1}{16\pi\mu_0} \frac{r_{1,m} r_{1,i} r_{1,k} }{r_1^5} S_{ik},
\end{align}
The desired disturbance vorticity is,
\begin{align}\label{eq:angularvelone}
\begin{split}
\omega_{p} &=\frac{1}{2} \epsilon_{pqm} \frac{\partial v_m}{\partial x_q} = \frac{1}{2}\frac{9a^3\mu_1}{16\pi\mu_0} S_{ik} \epsilon_{pqm} \left[ \frac{\delta_{qm} r_{1,i} r_{1,k} + r_{1,m} \delta_{qi} r_{1,k} + r_{1,m} r_{1,i} \delta_{qk}}{r_1^5} - 5 \frac{r_{1,m} r_{1,i} r_{1,k} r_{1,q}}{r_1^5} \right] \\
&= \frac{9a^3\mu_1}{16\pi\mu_0} \frac{1}{2d^3} \hat{d}_m \hat{d}_k \epsilon_{pim} [S_{ik} + S_{ki}] =  -\frac{9a^3\mu_1}{16\pi\mu_0} \frac{\epsilon_{pmj} \hat{d}_m U_j}{d^7}.
\end{split}
\end{align}
We next find the net vorticity due to multiple point viscosities arranged in a hemispherical shell. Substituting, $\hat{\bm{d}}=[\sin\theta \cos\phi, \sin\theta \sin \phi, \cos\theta]$ and $\hat{\bm{U}}=[\sin\Theta , 0, -\cos\Theta]$ ($\Theta$ is simply a reference angle to track how the velocity changes in time),
\begin{align}
\bm{\omega} = -\frac{9a^3\mu_1}{16\pi\mu_0}  \frac{1}{d_\alpha^7} [-\sin\theta \sin \phi \cos\Theta, \cos\theta \sin\Theta + \sin\theta \cos\phi \cos\Theta, - \sin\theta \sin \phi \sin\Theta ]
\end{align}
Summing over $\phi$, yields $\sum_{k_\phi = 0}^{N_\phi-1}  \cos 2 \pi k_\phi/N_\phi = 0$ and $\sum_{k_\phi = 0}^{N_\phi-1}  \sin 2 \pi k_\phi/N_\phi = 0$ resulting in $\omega_x =  \omega_z = 0$. The only non-zero term is,
\begin{align}
\omega_y &=   -\frac{9a^3\mu_1}{16\pi\mu_0}  \frac{1}{d_\alpha^7} \sin\Theta \sum_{k_\phi = 0}^{N_\phi-1} \sum_{k_\theta= 1}^{N_\theta - 1} \cos (k_\theta \pi/2 N_\theta)   = -\frac{9a^3\mu_1}{16\pi\mu_0}  \frac{1}{d_\alpha^7} \frac{N_\phi}{2} \left[\cot \left( \frac{\pi}{4N_\theta} \right) - 1 \right] \sin\Theta
\end{align}
\begin{figure}
    \centering
    \includegraphics[width=0.5\linewidth]{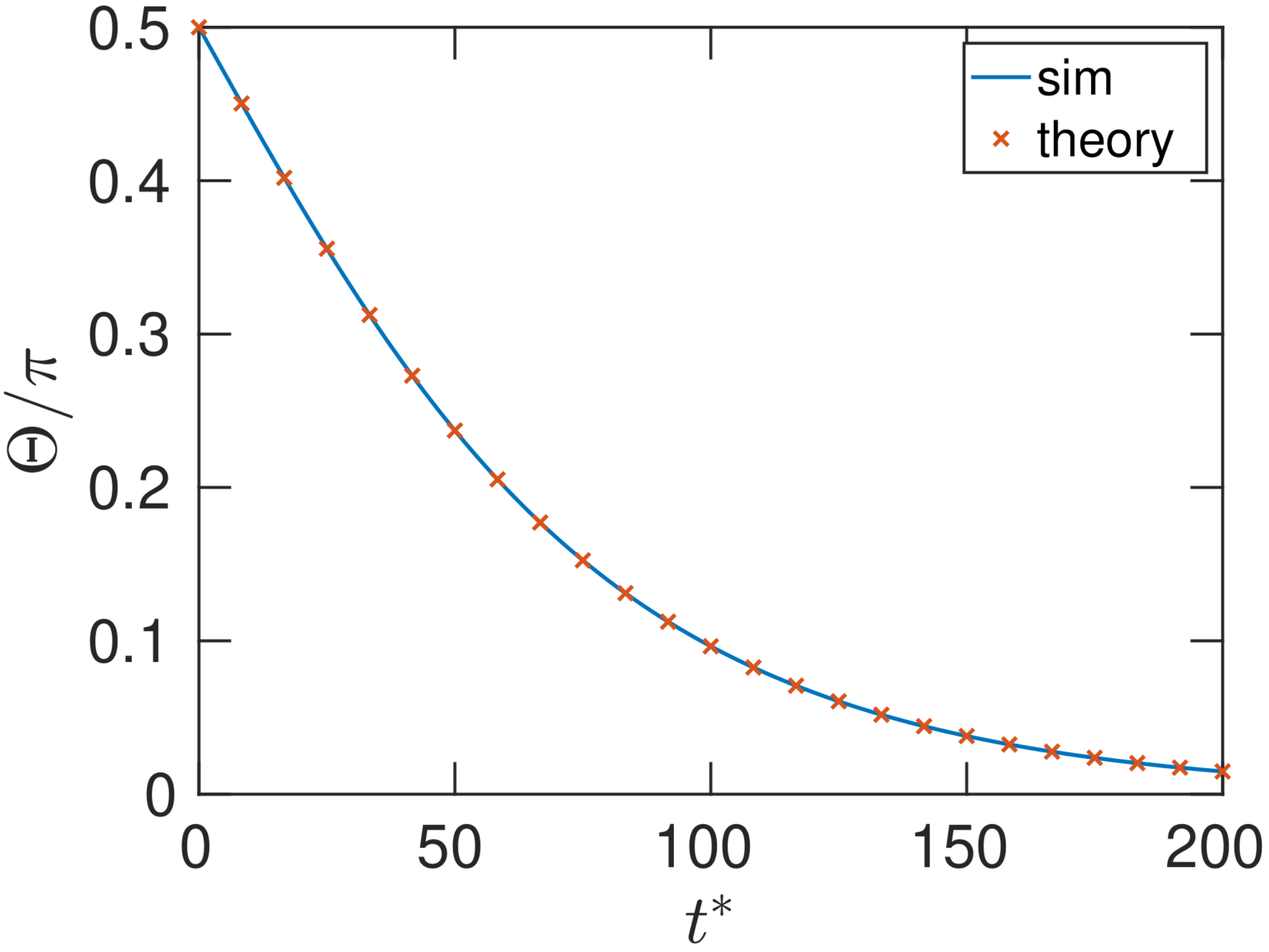}
    \caption{Angle, $\Theta$, scaled with $\pi$, made by the translational velocity of a spherical particle placed next to 57 viscosity sinks at a distance $d_\alpha^*=1.2$ and the effective viscosity gradient, plotted as a function of dimensionless time, $t^*=tU/a$. Comparison of Eq.~\eqref{angveltheory} (red cross marks) and numerically integrating  Eq.~\eqref{eq:angularvelone} (blue line) obtained by summing the effect of all viscosity sinks and rotating the translational velocity vector, $\bm{U}$, with $\bm{\Omega}$ is shown. All the viscosity sinks are identical in strength, $\mu_\alpha = -0.01\mu_0$.}
    \label{fig1sup}
\end{figure}
Note that we have not added any sinks around the equator of the particle by choice as it does not affect the results. We then add one sink at the top, i.e.~for $\theta=\pi/2$, to obtain the desired expression for the angular velocity,
\begin{align}
\Omega_y &= \omega_y = - \beta \sin \Theta, \quad \text{where}~~\beta = \frac{9a^3\mu_1}{16\pi\mu_0 d_\alpha^7}  \left[ 1 + \frac{N_\phi}{2} \left\{ \cot \left( \frac{\pi}{4N_\theta} \right) - 1 \right\} \right]
\end{align}
Next, we can integrate this equation in time, $\partial_t \Theta  = -\Omega_y$, to obtain an expression for $\Theta (t)$,
\begin{align}\label{angveltheory}
\Theta(t) = 2 \tan^{-1} \left[\tan \left\{\frac{\Theta (0)}{2}  \right\} \exp{(\beta t)}\right].
\end{align}

\section{Boundary Integral Equation}\label{sec:appendixD}

Hydrodynamics of an arbitrarily shaped rigid particle in an arbitrarily varying viscosity field in space is given by the boundary integral equation,
\begin{align}
\begin{split} 
\bm{u}(\bm{x}_0) &= -\frac{1}{8\pi\mu_0} \iint_S [\bm{f}(\bm{x}) \bcdot \mathsfbi{G}(\bm{x},\bm{x}_0) - \mu(\bm{x}) \bm{u}(\bm{x}) \bcdot \mathsfbi{T}(\bm{x},\bm{x}_0) \bcdot \bm{n}(\bm{x})] \, \mathrm{d}S(\bm{x}) \\
& + \frac{1}{8\pi\mu_0} \iiint_V \bnabla \mu(\bm{x}) \bcdot \mathsfbi{T}(\bm{x},\bm{x}_0) \bcdot \bm{u}(\bm{x})] \, \mathrm{d}V(\bm{x}),
\end{split} 
\end{align}
where $\bm{G}$ is the free-space Green's function for Stokes equation called Stokeslet and $\mathsfbi{T}$ is the corresponding stress tensor, called stresslet,
\begin{align}
\mathsfbi{G} = \frac{\bm{I}}{r_0} + \frac{\bm{r}_0 \bm{r}_0 }{r_0^3}, \quad
\mathsfbi{T}= -6\frac{ \bm{r}_0  \bm{r}_0  \bm{r}_0}{r_0^5}.
\end{align}
For rigid body motion, $\bm{u}(\bm{x}_0) = \bm{U} + \bm{\Omega} \btimes \bm{x}_0$, the double layer integral vanishes. Substituting, $\mu(\bm{x}) = \mu_0 + \mu_\alpha  \delta (\bm{x}-\bm{x}_\alpha)$, the boundary integral equation simplifies to,
\begin{align}
\begin{split} 
\bm{u}(\bm{x}_0) &= -\frac{1}{8\pi\mu_0} \iint_S \bm{f}(\bm{x}) \bcdot \mathsfbi{G}(\bm{x},\bm{x}_0) \, \mathrm{d}S(\bm{x}) - \frac{ \mu_\alpha}{8\pi\mu_0} \mathsfbi{T}(\bm{x}_\alpha,\bm{x}_0) \bm{:} \bnabla \bm{u}(\bm{y}) \Big|_{\bm{y}=\bm{x}_\alpha}
\end{split} 
\end{align}
If $\bm{x}_0 = \bm{x}_\alpha$, we do not consider the effect of the flow created by the point viscosity on its own location. The second term involving the gradient of the velocity field can be written as,
\begin{align}
\begin{split} 
-\frac{\mu_\alpha}{8\pi\mu_0} \mathsfbi{T}(\bm{x}_\alpha,\bm{x}_0) \bm{:} \bnabla_y\bm{u}(\bm{y})\Big|_{\bm{y}=\bm{x}_\alpha} &= \frac{1}{8\pi\mu_0} \frac{\mu_\alpha}{8\pi\mu_0} \mathsfbi{T}(\bm{x}_\alpha,\bm{x}_0) \bm{:}  \iint_S \bm{f}(\bm{x}) \bcdot  \bnabla_y \mathsfbi{G}(\bm{x},\bm{y})\Big|_{\bm{y}=\bm{x}_\alpha}  \, \mathrm{d}S(\bm{x}) 
\end{split} 
\end{align}
Substituting this back in the integral equation, we get,
\begin{align}
\begin{split} 
u_j(\bm{x}_0) &= -\frac{1}{8\pi\mu_0}\left [ \iint_S f_l(\bm{x}) \left( G_{lj}(\bm{x},\bm{x}_0)  - \frac{\mu_\alpha}{8\pi\mu_0} T_{ijk}(\bm{x}_\alpha,\bm{x}_0) \bnabla_{y,i} G_{lk}(\bm{x},\bm{y})\Big|_{\bm{y}=\bm{x}_\alpha}  \right) \, \mathrm{d}S(\bm{x}) \right]
\end{split} 
\end{align}
Next, we evaluate the gradient of the Green's function,
\begin{align}
\begin{split}
\bnabla_{y,i} G_{lk}(\bm{x},\bm{y})\Big|_{\bm{y}=\bm{x}_\alpha} &= \frac{\partial }{\partial y_i}\left( \frac{\delta_{lk}}{r} + \frac{r_lr_k}{r^3} \right) = -\frac{\partial }{\partial r_i}\left( \frac{\delta_{lk}}{r} + \frac{r_lr_k}{r^3} \right) \\
&= -\frac{\partial }{\partial r_i}\left( \frac{\delta_{lk}}{r} + \frac{r_lr_k}{r^3} \right) =  \frac{[\delta_{lk}r_i - \delta_{li} r_k - r_l\delta_{ik}]r^2 + 3r_ir_lr_k}{r^5} 
\end{split} 
\end{align}
where $\bm{r} = \bm{x} - \bm{x}_\alpha, ~~ \bm{x} \in S$.  

The modification to the Green's function is found by substituting $\bm{p}  = \bm{x}_\alpha - \bm{x}_0$,
\begin{align}
G_{v,lj}(\bm{x},\bm{x}_0,\bm{x}_\alpha) =&  \frac{\mu_\alpha}{8\pi\mu_0} T_{ijk}(\bm{x}_\alpha,\bm{x}_0)\bnabla_{y,i} G_{lk}(\bm{x},\bm{y})\Big|_{\bm{y}=\bm{x}_\alpha} = \frac{3\mu_\alpha}{4\pi\mu_0} \frac{r_l p_j}{p^5r^5} [p^2 r^2 - 3(\bm{p} \bcdot \bm{r})^2]
\end{align}

\bibliography{papers} 

\end{document}